\documentclass[11pt,twoside]{article}
\usepackage{asp2010}
\usepackage{graphicx}

\resetcounters 

\begin{document}

\resetcounters

\title{Extreme-mass-ratio bursts from the Galactic Centre}
 \author{Christopher~P.~L.~Berry and Jonathan R.\ Gair \affil{Institute of Astronomy, University of Cambridge, Cambridge, UK}}
 
\begin{abstract} 
An extreme-mass-ratio burst (EMRB) is a gravitational wave signal emitted when a compact object passes through periapsis on a highly eccentric orbit about a much more massive body, in our case a stellar mass object about the $4.31 \times 10^6 M_\odot$ massive black hole (MBH) in the Galactic Centre. We investigate how EMRBs could constrain the parameters of the Galaxy's MBH. EMRBs should be detectable if the periapsis is $r_\mathrm{p} < 65 r_\mathrm{g}$ for a $\mu = 10 M_\odot$ orbiting object, where $r_\mathrm{g} = GM_\bullet/c^2$ is the gravitational radius. The signal-to-noise ratio $\rho$ scales like $\log(\rho) \simeq -2.7\log(r_\mathrm{p}/r_\mathrm{g}) + \log(\mu/M_\odot) + 4.9$. For periapses smaller than $\sim 10 r_\mathrm{g}$, EMRBs can be informative, providing good constraints on both the MBH's mass and spin.
\end{abstract}

\section{Introduction} 

We believe that most galactic nuclei have harboured a massive black hole (MBH) during their evolution~\citep{Lynden-Bell1971, Rees1984}. Observations show there are correlations between the MBHs' masses and their host galaxies' properties~\citep{Kormendy1995, Magorrian1998, Graham2011}. These suggest coeval evolution of MBH and galaxy~\citep{Peng2007, Jahnke2011}. Since they share a common history; one can inform us of the other.

The best opportunity to study MBHs comes from the compact object in our own galactic centre (GC), coincident with Sagittarius A* (Sgr A*). This is an MBH of mass $M_\bullet = 4.31 \times 10^6 M_\odot$ at a distance of only $R_0 = 8.33~\mathrm{kpc}$~\citep{Gillessen2009}.

According to the no-hair theorem, the MBH should be described completely by its mass $M_\bullet$ and (dimensionless) spin $a_\ast$~\citep{Chandrasekhar1998}. The spin is related to the BH's angular momentum $J$ by $a_\ast = cJ/GM_\bullet^2$. As we have a good estimate of the mass, to gain a complete description we have only to measure the spin.

An MBH accumulates mass and angular momentum through accretion~\citep{Volonteri2010}. There are several possible accretion mechanisms, each leaving its own imprint on the final spin. Measuring the spin of the MBH shall help us understand the relative importance of these processes, and gain insight into the Galaxy's past.

An exciting means of inferring information about the MBH is through gravitational waves (GWs) emitted when stellar mass compact objects (COs) pass close by~\citep{Sathyaprakash2009}. A space-borne detector can detect GWs in the frequency range of interest for these encounters~\citep{Danzmann2003, Amaro-Seoane2012a}. The waveforms generated when COs inspiral towards an MBH have been much studied~\citep{Glampedakis2005}. These systems are typically formed following two-body encounters so the initial orbits are highly eccentric; a burst of radiation is emitted during each periapse passage. These are extreme mass-ratio bursts (EMRBs; \citealt*{Rubbo2006}). These orbits can evolve, becoming more circular, and then emitting continuously in the detector frequency range. These signals are extreme mass-ratio inspirals (EMRIs; \citealt{Amaro-Seoane2007}).

We investigate the bursts from the initial highly-eccentric orbits, under the simplifying assumption that these orbits are marginally bound, or parabolic, since highly eccentric orbits appear almost indistinguishable from an appropriate parabolic orbit.\footnote{Here ``parabolic'' and ``eccentricity'' refer to the energy of the geodesic and not to its geometric shape.} The event rate for detectable EMRBs has been estimated as $1~\mathrm{yr^{-1}}$~\citep*{Hopman2007}. Even if only a single burst is detected, this should still give an unparalleled probe of the spacetime of the GC. What can be inferred depends upon the orbit, which we investigate here. 

We use the classic \textit{Laser Interferometer Space Antenna} (\textit{LISA}) design for this work. It is hoped that any future detector shall have comparable sensitivity to \textit{LISA}, and that studies using this design shall be a sensible benchmark.

\section{Numerical kludge waveforms}

For given angular momenta, and initial starting position, we can calculate the geodesic trajectory in a Kerr background. The orbiting body is assumed to follow this track exactly; we ignore evolution due to the radiation, which is negligible for EMRBs. From this trajectory we calculate the waveform using a semirelativistic approximation~\citep{Ruffini1981}: we assume the particle moves along a geodesic, but radiates as in flat spacetime. This is known as a numerical kludge (NK), and well approximates results computed by more accurate methods~\citep{Babak2007}. The use of the geodesic ensures the correct frequency components appear in the waveform, but the flat-space wave generation means they do not have precisely the correct amplitudes.

NK approximations aim to encapsulate the main characteristics of a waveform by using the exact particle trajectory (ignoring inaccuracies from radiative effects and from the particle's self-force), whilst saving on computational time by using approximate waveform generation.

We build an equivalent flat-space trajectory from the Kerr geodesic. This is done by identifying the Boyer-Lindquist coordinates~\citep{Boyer1967} with a set of flat-space coordinates. We use spherical polars so $\{r_\mathrm{BL},$ $\theta_\mathrm{BL},$ $\phi_\mathrm{BL}\} \rightarrow \{r_\mathrm{sph}, \theta_\mathrm{sph}, \phi_\mathrm{sph}\}$ \citep*{Gair2005}. Oblate-spheroidal coordinates yield similar results.

With a flat-space trajectory, we may use a flat-space wave generation formula: the quadrupole-octupole formula~\citep{Bekenstein1973, Press1977, Yunes2008}. This is correct for a slowly moving source, and is the familiar quadrupole formula, derived from linearized theory, plus the next order terms.  

\section{Waveforms and detectability}\label{sec:Waveforms}

\subsection{Model parameters}

The waveform depends on the properties of the MBH; the CO and its orbit, and the detector. We assume the position of the detector is known, and the MBH is coincident with the radio source of Sgr A*, which is within $20 r_\mathrm{g}$ of the MBH~\citep{Reid2003,Doeleman2008}. We use the J2000.0 coordinates, which are determined to high accuracy~\citep{Reid1999, Yusef-Zadeh1999}. The parameters left to infer are: (1) The MBH's mass $M_\bullet$. This is well constrained by the observation of stellar orbits about Sgr A*~\citep{Ghez2008, Gillessen2009}, the best estimate is $M_\bullet = (4.31 \pm 0.36) \times 10^6 M_\odot$. (2) The spin parameter $a_\ast$. (3, 4) The orientation angles for the black hole spin $\Theta_\mathrm{K}$ and $\Phi_\mathrm{K}$. (5) The ratio of the GC distance and the compact object mass $\zeta = R_0/\mu$. This scales the amplitude of the waveform. Bursts do not undergo orbital evolution, hence we cannot break the degeneracy between $R_0$ and $\mu$. The distance is constrained by stellar orbits to be $R_0 = 8.33 \pm 0.35~\mathrm{kpc}$~\citep{Gillessen2009}. (6, 7) The angular momentum of the CO, parametrized by the magnitude at infinity $L_\infty$ and the orbital inclination $\iota$. (8--10) Coordinates specifying the trajectory. We use the angular phases at periapse, $\phi_\mathrm{p}$ and $\theta_\mathrm{p}$, as well as the time of periapse $t_\mathrm{p}$.

\subsection{Signal-to-noise ratio}

The detectability of a burst depends upon its SNR $\rho$. The amplitude of the waveform is proportional to the CO mass; we work in terms of a mass-normalised SNR $\hat{\rho} = (\mu/M_\odot)^{-1}\rho$. We considered a range of orbits. The spin of the MBH and the orbital inclination were randomly chosen, and the periapse distance was set so that the distribution would be uniform in log-space (down to the inner-most orbit). For each set of the extrinsic parameters, the periapse positions, orientation of the MBH, and orbital position of the detector were varied: we used five combinations of these intrinsic parameters (each drawn from a separate uniform distribution), taking the mean of $\ln \rho$ for each set.

The correlation between the periapse radius and SNR is shown in fig.~\ref{fig:SNR}.
\articlefigure[width=0.5\textwidth]{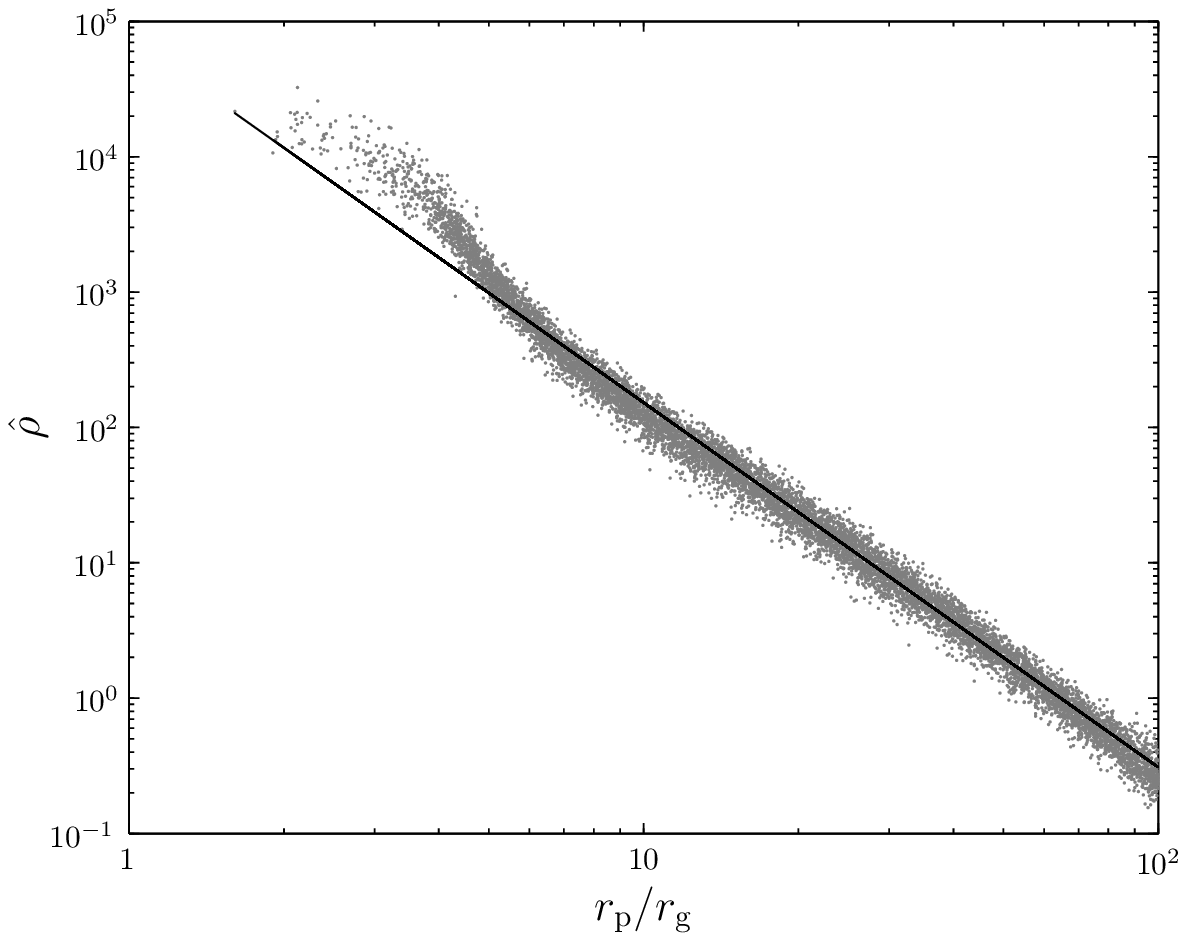}{fig:SNR}{Mass-normalised SNR as a function of periapse radius. The points are the values averaging over each set of intrinsic parameters. The best fit line is $\log(\hat{\rho}) = -2.69\log(r_\mathrm{p}/r_\mathrm{g}) + 4.88$ for orbits with $r_\mathrm{p} >  13.0 r_\mathrm{g}$.}
The shape is predominantly determined by the noise curve. The change in the trend reflects the transition from approximately power law behaviour to the bucket of the noise curve. We fit a fiducial power law to orbits with a characteristic frequency of $f_\ast = \sqrt{GM_\bullet/r_\mathrm{p}} < 1 \times 10^{-3}~\mathrm{Hz}$, to avoid spilling into the bucket. Changing the cut-off within a plausible region alters the fit coefficients by $\sim 0.1$.

Setting a threshold of $\rho = 10$, a $1 M_\odot$ ($10 M_\odot$) CO is detectable if $r_\mathrm{p} < 27 r_\mathrm{g}$ ($65 r_\mathrm{g}$).

\section{Parameter estimation \& results}\label{sec:Estimation}

Having detected a GW signal, we are interested in what we can learn about the source. We performed Markov chain Monte Carlo (MCMC) simulations to characterise the posterior probability distribution~\citep[chapter 29]{MacKay2003}. Waveforms were computed for a range of orbits. In each case the MBH has the standard mass and position. The CO was chosen to be $10 M_\odot$, as the most promising candidates for EMRBs would be stellar mass black holes: they are massive and hence produce higher SNR bursts, they are more likely to be on close orbits as a consequence of mass segregation, and they cannot be tidally disrupted. Orbits were chosen with periapses uniformly distributed in logarithmic space (to the inner-most orbit). The other parameters were chosen randomly from appropriate uniform distributions. 

It is possible to place good constraints from the closest orbits, although there is significant correlation between parameters. The standard deviation $\sigma$ of the recovered posteriors are shown in fig.~\ref{fig:sigmas} for the mass and spin. Filled circles are used for converged MCMC runs. Open circles are for those yet to converge; widths should be accurate to a factor of a few.
\articlefiguretwo{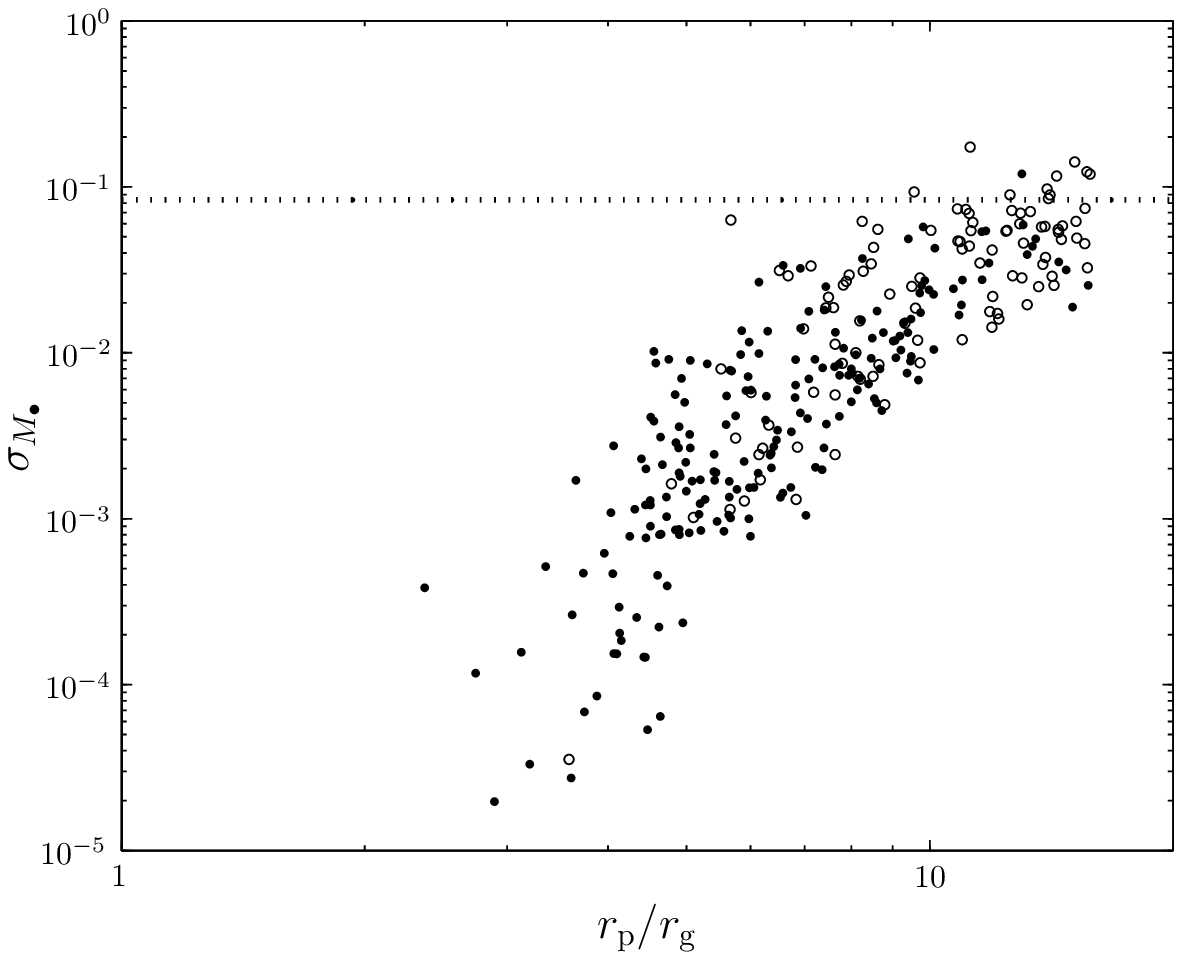}{Fig_MCMC_SD_rp_2}{fig:sigmas}{Distribution widths as functions of periapse $r_\mathrm{p}$. Filled circles are used for converged runs; open circles for unconverged runs. The dotted line is the current uncertainty for $M_\bullet$; the dashed line the standard deviation for an uninformative prior, and the solid line the total prior range.}
These results do not incorporate any priors (save to keep them within realistic ranges). Therefore, the resulting distributions characterise what we could learn from EMRB's alone.

\subsection{Scientific potential}

The current uncertainty in the mass is $\sigma_{M_\bullet} = 0.36 \times 10^6 M_\odot$ ($\sim 8\%$). It appears that orbits of a $\mu = 10 M_\odot$ CO with periapses $r_\mathrm{p} \lesssim 13 r_\mathrm{g}$ should be able to match this. Accuracy of $1\%$ is possible if $r_\mathrm{p} \lesssim 8 r_\mathrm{g}$.

The spin is less well constrained. To obtain an uncertainty for the magnitude of $0.1$, comparable to that achieved in X-ray measurements of active galactic nuclei, the periapsis needs to be $r_\mathrm{p} \lesssim 11 r_\mathrm{g}$. As the spin encodes information of the merger and accretion history, this could illuminate the MBH's formation.

We have no {\it a priori} knowledge about the CO or its orbit, so anything we learn would be new. However, this is not particularly useful information, unless we observe multiple bursts, and can start to build up statistics for the dynamics of the GC. Using current observations for the distance to the GC, which could be further improved by the mass measurement from the EMRB, it is possible to infer a value for the mass $\mu$ from $\zeta$. This could inform us of the nature of the object (BH, NS or WD) and be a useful consistency check. A small value of $\zeta$, indicating a massive CO, would be unambiguous evidence for the existence of a stellar mass black hole.

\bibliography{EMRB}
\bibliographystyle{asp2010}

\end{document}